\documentclass[aps,prb,showpacs,preprint,superscriptaddress]{revtex4}
\usepackage{graphicx}
\usepackage{amssymb,amsmath} 

\begin{document}

\preprint{}

\title{Spin-polarization of platinum~(111) induced by the proximity to cobalt nanostripes}

\author{Focko Meier}
\affiliation{Institute of Applied Physics, Hamburg University, Jungiusstrasse 11, D-20355 Hamburg, Germany}
\author{Samir Lounis}
\affiliation{Department of Physics and Astronomy, University of California Irvine, California, 92697 USA}
\author{Jens Wiebe}
\affiliation{Institute of Applied Physics, Hamburg University, Jungiusstrasse 11, D-20355 Hamburg, Germany}
\author{Lihui Zhou}
\affiliation{Institute of Applied Physics, Hamburg University, Jungiusstrasse 11, D-20355 Hamburg, Germany}
\author{Swantje Heers}
\affiliation{Institut f\"ur
Festk\"orperforschung \& Institute for Advanced Simulation, Forschungszentrum J\"ulich \& JARA, D-52425 J\"ulich,
Germany}
\author{Phivos Mavropoulos}
\affiliation{Institut f\"ur
Festk\"orperforschung \& Institute for Advanced Simulation, Forschungszentrum J\"ulich \& JARA, D-52425 J\"ulich,
Germany}
\author{Peter H. Dederichs}
\affiliation{Institut f\"ur
Festk\"orperforschung \& Institute for Advanced Simulation, Forschungszentrum J\"ulich \& JARA, D-52425 J\"ulich,
Germany}
\author{Stefan Bl\"ugel}
\affiliation{Institut f\"ur
Festk\"orperforschung \& Institute for Advanced Simulation, Forschungszentrum J\"ulich \& JARA, D-52425 J\"ulich,
Germany}
\author{Roland Wiesendanger}
\affiliation{Institute of Applied Physics, Hamburg University, Jungiusstrasse 11, D-20355 Hamburg, Germany}

\date{\today}

\begin{abstract}
We measured a spin polarization above a Pt (111) surface in the vicinity of a Co nanostripe by spin-polarized scanning tunneling spectroscopy. The spin polarization is exponentially decaying away from the Pt/Co interface and is detectable at distances larger than 1~nm. By performing self-consistent ab-initio calculations of the electronic-structure for a related model system we reveal the interplay between the induced magnetic moments within the Pt surface and the spin-resolved electronic density of states above the surface.
\end{abstract}

\maketitle

\section{Introduction}
The remarkable properties of magnetic nanostructures grown on non-magnetic metal substrates rely significantly on the electronic coupling between the atoms within the nanostructure and substrate atoms underneath.~\cite{bluegel_prl_1988} This electronic coupling determines e.g. the strength and direction of the magnetic anisotropy as well as the total magnetic moment.~\cite{gambardella2003} Additionally the substrate electrons govern the collective behavior of ensembles of magnetic nanostructures, e.g. by providing ferromagnetic order due to indirect exchange interaction between separated magnetic nanostructures.~\cite{pierce_prl_2004} This interaction, also known as Ruderman-Kittel-Kasuya-Yosida (RKKY) interaction, has been found in diluted magnetic systems, where magnetic 3d impurity atoms are dissolved in non-magnetic host metals.~\cite{RK1954,K1956,Y1957} In these samples, the localized magnetic moment of an impurity atom is screened by a spatially oscillating long range spin-polarization of the host conduction electrons.~\cite{Graham_PRL_1966} Therefore the distance between impurity atoms determines the sign and strength of the interaction, respectively. The same coupling has recently been observed directly for atoms \textit{on} surfaces.~\cite{Meier_science_2008,zhou_np_2010} A second important effect takes place for magnetic 3d impurity atoms in host metals which nearly fulfill the Stoner criterion, such as Pt and Pd, i.e. they are nearly ferromagnetic and are therefore characterized by a high susceptibility. In these so called giant-moment dilute alloys the 3d impurities induce relatively strong magnetic moments in the neighboring host atoms which form a spin-polarized cluster.~\cite{crangle_1960} Since this effect can cause an additional exchange interaction between magnetic atoms in nanostructures it is important to obtain knowledge about the size of the polarization cloud and the decay of the induced magnetization with increasing distance from the magnetic atom.~\cite{skomski_2008,oswald_prl_1986}\\
Both mechanisms are considered to be important for multilayer systems~\cite{parkin_prl_1990}, like Co-Pt, which consist of sequences of ferromagnetic Co layers separated by non-magnetic Pt spacer layers.~\cite{bruno_prl_1992, dang_prb_1994} The magnetic interlayer coupling between the ferromagnetic layers often shows deviations from a pure RKKY behavior, indicating that other mechanisms contribute to the total magnetic interaction. One contribution originates from magnetoelastic interactions due to interface roughness between the magnetic and non-magnetic layers \cite{moritz_europhys_2004,Li_solid_state_2008} while with decreasing temperatures the induced magnetic moments of Pt become relevant for the magnetic coupling.~\cite{knepper_prb_2005} In order to qualify specific contributions to the overall interaction a profound knowledge on the local configuration of the interface is required.\\
In this work we present a combined experimental and theoretical study on the spin-polarization of Pt in the vicinity of Co nanostripes on a Pt(111) surface. We use spin-resolved scanning tunnelling spectroscopy~\cite{wiesendanger_RevModPhys_2009} and the Korringa-Kohn-Rostoker Green function method (KKR) within the framework of density functional theory.~\cite{skkr} Our experimental technique allows to obtain an extensive knowledge concerning the topographic, electronic as well as magnetic properties of the sample. We show that the measured Pt \textit{local electronic density of states}~(LDOS) near the Fermi energy in the vacuum exhibits an exponentially decaying spin-polarization indicating magnetic moments induced by the Co nanostripe. Interestingly this effect can be observed for lateral distances from the Co nanostripe larger than four Pt lattice spacings where the RKKY interaction provides already an antiferromagnetic coupling as shown in a previous study.~\cite{Meier_science_2008} 
The calculated induced magnetic moments in the Pt surface close to embedded Co atoms show a distance dependent oscillation between ferromagnetic and antiferromagnetic alignment, while the vacuum spin-polarization at particular energies experiences an exponential decay in the lateral direction.

\section{Experimental setup}
All experiments were performed in an ultrahigh-vacuum system containing a home-built 300~mK STM operating at a magnetic field $B$ up to 12~$T$ perpendicular to the sample surface.~\cite{wiebe_rsi_2004} In this work we used Cr-covered W tips, which are sensitive to the out-of-plane direction of the nanostripe magnetization $\vec{M}_{Co}$.~\cite{kubetzka2002,boderep} In order to retain a strong spin polarization the tips were eventually dipped into Co stripes.~\cite{Meier_science_2008,meier_prb_2006} This procedure can result in attaching Co clusters to the tip apex which affects the magnetic $B$~field required to switch the tip magnetization $\vec{M}_{tip}$. Further details on the sample and tip preparation are given in Refs.~\cite{Meier_science_2008,meier_prb_2006}. Co was evaporated at two different temperatures on a clean Pt(111) crystal. First, a tenth of an atomic layer (AL) was deposited at room temperature leading to Co nanostripes attached to the Pt(111) step edges. At a temperature below 25 K a much smaller amount was evaporated which resulted in a tiny number of single Co adatoms randomly distributed on the surface.

\section{Experimental results}
\begin{figure}
\includegraphics[width=0.7\columnwidth]{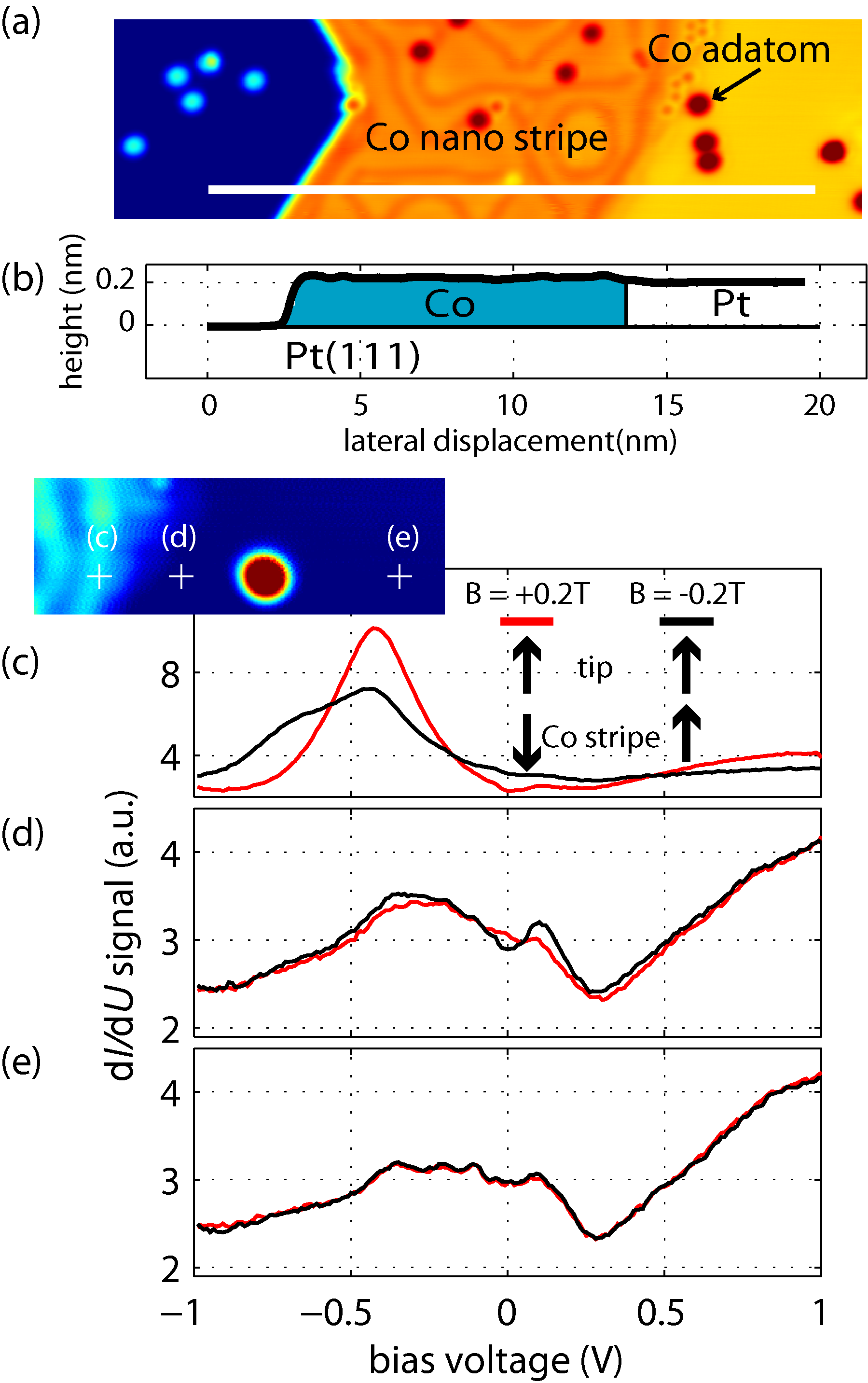}
\caption{~(a)~STM~topograph of two Pt(111) terraces with individual Co adatoms and a Co nanostripe attached to a Pt step edge. (b)~Line section along the line marked in (a).~(c-e) d$I$/d$U$ spectra taken at positions given in inset, which displays the interface between the Co nanostripe (left) and the Pt terrace (right). The relative orientations of tip and Co nanostripe magnetization,~$\vec{M}_{tip}$ and $\vec{M}_{Co}$, are indicated by arrows.~(Tunnelling parameters: $U_{stab}$=1.0~V, lock-in modulation voltage $U_{mod}$=10~mV, stabilization current $I_{stab}$=1.0~nA, $T=$0.3~K) \label{fig:1}}
\end{figure}

Figure~\ref{fig:1}~(a) shows a Co nanostripe attached to a Pt step edge between two Pt terraces and individual Co adatoms. The one AL high Co nanostripe can be easily identified by a dense network of dislocation lines originating from the lattice mismatch between Co and Pt.~\cite{grutter1994,lundgren2000,meier_prb_2006} Obviously the Co stripe appears 20~pm higher than the Pt as visible in the line section in Fig.~\ref{fig:1}~(b). Information regarding the spin-resolved LDOS in the vacuum above the Co nanostripe as well as above the Pt surface is obtained by measuring the differential conductance d$I$/d$U$ as a function of location ${\bf r}$, the applied bias voltage $U_{stab}$ as well as the relative orientation between tip magnetization $\vec{M}_{tip}$ and the sample magnetization $\vec{M}_{Co}$.~\cite{Wortmann_PRL_2001} From previous measurements on the nanostripes it is known that $\vec{M}_{Co}$ is oriented out-of-plane.~\cite{meier_prb_2006} 

\begin{figure}
\includegraphics[width=0.7\columnwidth]{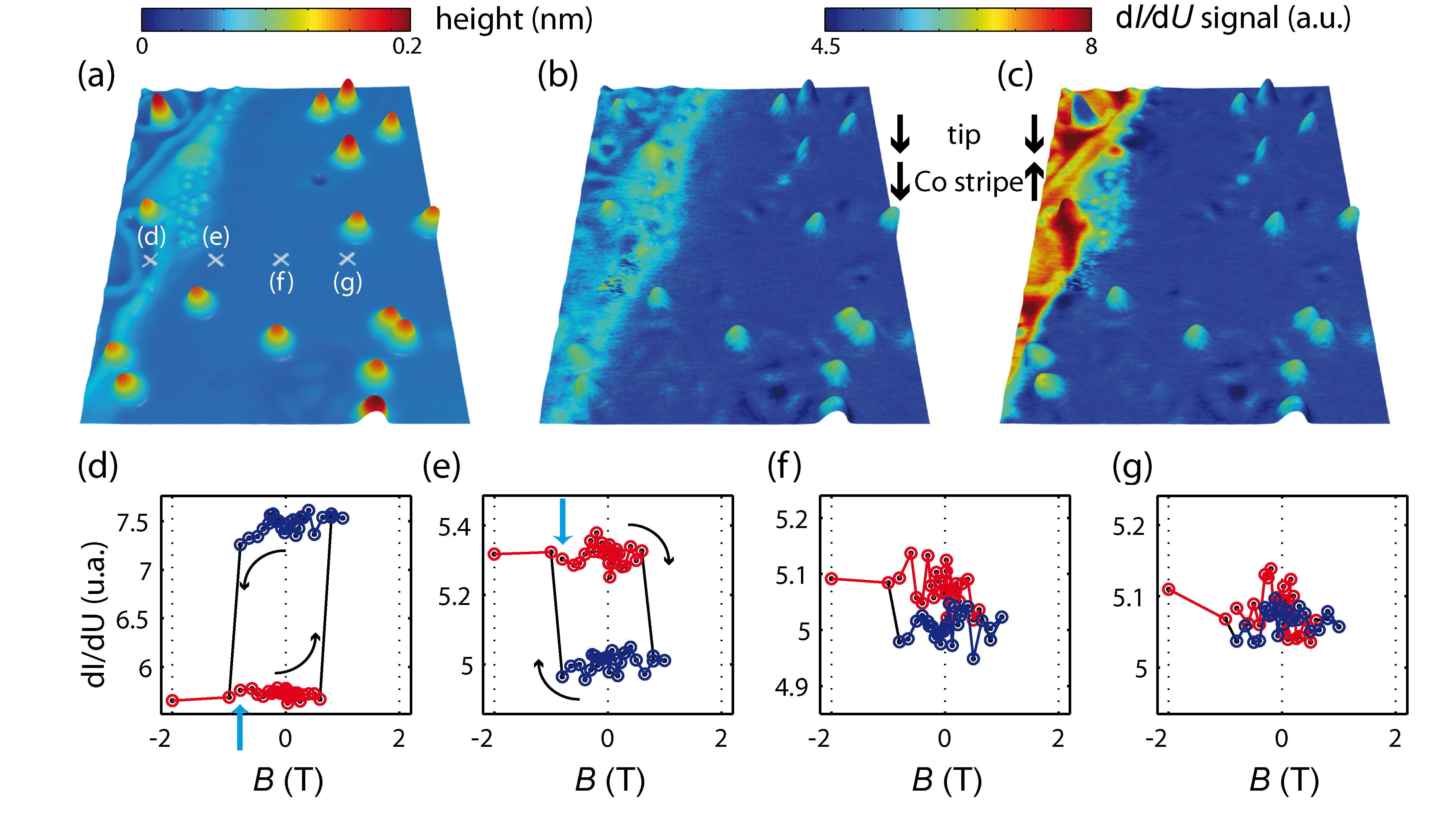}
\caption{~(a)~STM topograph in 3D view (size $11.6 \times 15.6$~nm$^2$).~(b),(c)~STM topograph in 3D view colorized with the simultaneously recorded spin-resolved d$I$/d$U$ map obtained at $B=+0.6$~T and $B=+1.0$~T, respectively. Relative orientation of $\vec{M}_{tip}$ and $\vec{M}_{Co}$ is indicated by arrows.~(Tunnelling parameters: $U_{stab}$=0.3~V, $U_{mod}$=20~mV, $I_{stab}$=0.8~nA) (d)-(g)~Magnetization curves taken at positions marked by crosses in Fig.\ref{fig:1}(a). Positions are separated by 2.3~nm. Arrows in (d) and (e) mark the start and direction of rotation of the $B$~field loop. Red and and color indicate $dI/dU$~values representing parallel and antiparallel orientation of $\vec{M}_{tip}$ and $\vec{M}_{Co}$ for each hysteresis.\label{fig:2}}
\end{figure}

\begin{figure}
\includegraphics[width=0.7\columnwidth]{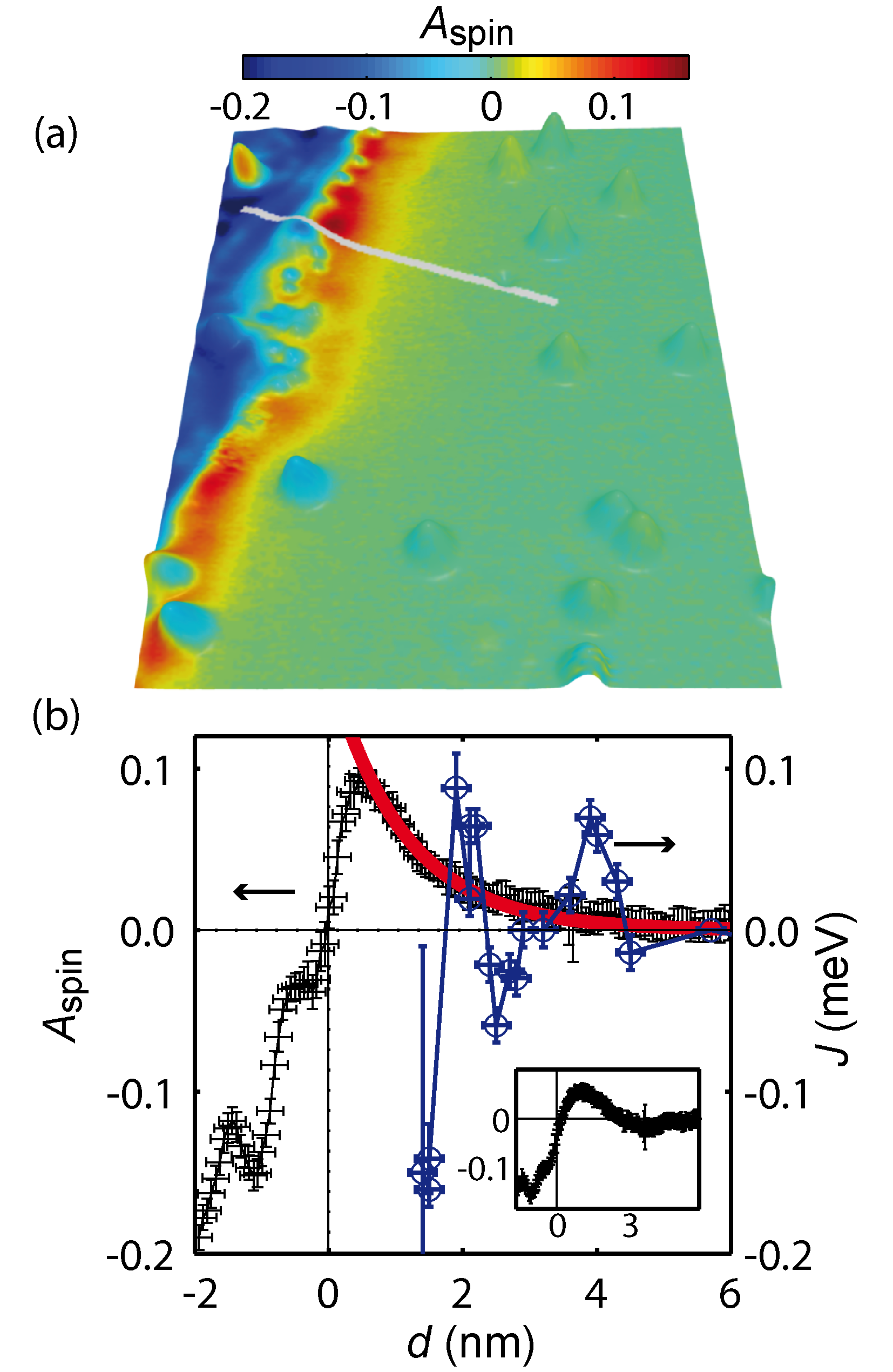}
\caption{~(a)~STM topograph in 3D view colorized with the calculated asymmetry map obtained from local magnetization curves~($U_{stab}$=0.3~V).~(b)~Crosses: asymmetry values below line section indicated in (a). Open circles:~magnetic exchange energy $J$ for the coupling between Co nanostripe and individual Co atoms taken from Ref.~\cite{Meier_science_2008}. '0' indicates the border between Co nanostripe and Pt layer. The red line shows an exponential fit according to Eq.~\ref{eq:1}. Inset: asymmetry values as in (a) for $U_{stab}$=-0.1~V.  \label{fig:3}}
\end{figure}

Figure~\ref{fig:1}~(c-e) show the resulting d$I$/d$U({\bf r},U)$ spectra taken on locations indicated in the inset on the Co nanostripe and on the Pt(111) close and far from the nanostripe. Here, $\vec{M}_{tip}$ is switched up or down by $B$~fields of $+0.2$~T and $-0.2$~T while $\vec{M}_{Co}$ is constant. This allows to measure the d$I$/d$U$ signal for parallel and antiparallel alignment of $\vec{M}_{tip}$ and $\vec{M}_{Co}$. On the Co nanostripe (c.f. Fig.~\ref{fig:1}~(c)) the spin resolved d$I$/d$U$ spectra show a dominant peak located at -0.4~eV below $E_F$ which originates from a $d$-like Co surface resonance of minority-spin character.~\cite{meier_prb_2006} The intensity of this state is changing considerably for parallel and antiparallel alignment of $\vec{M}_{tip}$ and $\vec{M}_{Co}$. In contrast to that, the spectra on the bare Pt far from the nanostripe in Fig.~\ref{fig:1}~(e) do not show the electronic signature of the $d$-like surface resonance but the onset of the unoccupied surface state at $eU=0.3$~eV is visible.~\cite{wiebe2005} Furthermore, no dependency on $\vec{M}_{tip}$ is found as expected for a non-magnetic material. Figure~\ref{fig:1}~(d) shows spectra which have been taken on Pt but only at a distance of around 1~nm with respect to the Co nanostripe. The spectra show the typical signature of a bare Pt(111) surface far from the Co stripe (c.f. Fig.~\ref{fig:1}~(e)). However, a clear dependency on the relative orientation of $\vec{M}_{tip}$ and $\vec{M}_{Co}$ is now observed in an energy range from -0.5~eV to +0.5~eV around $E_F$. Neither from our topographic nor spectroscopic data we have any indications of Co incorporation into the Pt surface or sub-surface layers within the probed area.~\cite{lundgren1999,quaas_prb_2004} This experimental result already proves a spin-polarization of the clean Pt(111) at a distance of more than three lattice spacings to the Co nanostripe. \\
In order to obtain information about this induced spin polarization we probed the spatially resolved d$I$/d$U$ signal (d$I$/d$U$ map) in a boundary area shown in Fig.~\ref{fig:2}~(a). For this area d$I$/d$U$ maps have been recorded at $U_{stab}=+0.3$~V in a complete B-field loop starting from $-0.8$~T to $+1.0$~T and back to $-2.0$~T. Figures~\ref{fig:2}~(b) and (c) show exemplary 3D topographs colorized with the simultaneously measured d$I$/d$U$~maps obtained at $B=+0.6$~T and $B=+1.0$~T, where the relative orientation of $\vec{M}_{tip}$ and $\vec{M}_{Co}$ has changed due to a \textit{B} field induced $\vec{M}_{Co}$ reversal. The d$I$/d$U$~signal above the Pt terrace appears the same in both figures. However, a difference in d$I$/d$U$ intensity above Pt close to the stripe is observed.\\
From the sequence of $B$~field depending d$I$/d$U$~maps \textit{local magnetization curves} are obtained by plotting the d$I$/d$U$~signal at one image point as a function of $B$. Figures~\ref{fig:2}~(d)-- (g) show local magnetization curves taken at positions as marked in Fig.~\ref{fig:2}~(a). The magnetization curve of the Co stripe in Fig.~\ref{fig:2}~(d) shows two magnetic states and a square-like hysteresis indicating its ferromagnetic state and a coercivity of $B_c=0.7 \pm 0.05$~T. Strikingly, the magnetization curves measured on the Pt in the vicinity of the Co nanostripe show that there is an explicit link between the magnetic state of the Co stripe and the spin polarization measured on the Pt. Similar magnetization curves have been recorded for each point of the area of Fig.~\ref{fig:2}~(a). From these magnetization curves the so-called spin asymmetry $A_{\rm spin}$ is calculated by 
\begin{eqnarray}
                    A_{\rm spin} & = & \frac{dI/dU_{\uparrow\uparrow}-dI/dU_{\uparrow\downarrow}}{dI/dU_{\uparrow\uparrow}+dI/dU_{\uparrow\downarrow}}.
\end{eqnarray}
which characterizes the square-like magnetization curves and is a measure for the spin-polarization at $eU$ in the vacuum.~\cite{boderep} $dI/dU_{\uparrow\uparrow}$ and $dI/dU_{\uparrow\downarrow}$ denote the averaged values from all red and blue data points in the magnetization curves (Fig.~\ref{fig:2}~(d)-(g)), i.e for parallel and antiparallel alignment of $\vec{M}_{tip}$ and $\vec{M}_{Co}$ in each curve. An asymmetry value is obtained for each image point. This results in an asymmetry map shown in Figure~\ref{fig:3}~(a). The Co stripe shows a strong negative $A_{\rm spin}$ while on the Pt terrace far from the stripe $A_{\rm spin}$ is zero. Above the Pt close to the Co stripe an area with positive $A_{\rm spin}$ is visible which fades out for an increasing distance from the nanostripe. The decay is further analyzed in Fig.~\ref{fig:3}~(b) which shows $A_{\rm spin}$ values below the section line in Fig.~\ref{fig:3}~(a) as a function of the distance $d$ from the Co nanostripe.
In order to quantify the decay behavior the graph in Fig.~\ref{fig:3}~(b) has been fitted to a simple exponential function 
\begin{eqnarray}
                    f=Ce^{-x/\lambda}
\label{eq:1}                    
\end{eqnarray} 
where $C$ and $\lambda$ denote the amplitude and the decay length, respectively. Even though the exact value of $\lambda$ depends on the specific line section, values in the range from $\lambda=0.9$~nm to $\lambda=1.2$~nm are obtained corresponding to more than three next nearest neighbor distances within the Pt lattice. We observe the same quantitative behavior in $A_{\rm spin}$ calculated from d$I$/d$U({\bf r})$~maps recorded at $U_{stab}$=-0.1~V (cp. inset Fig.\ref{fig:3}(b)). Together with the dependency on the spin-resolved d$I$/d$U$-curves measured close to the Co stripe (cp. Fig.~\ref{fig:1}~(d)) we conclude that the observed spin-polarization above Pt is present in a large energy window around the Fermi energy. This result suggests that the measured spin-polarization is due to an exponentially decaying magnetic moment $\vec{M}_{Pt}$ induced by the vicinity to the Co nanostripe. 

Figure~\ref{fig:3}~(b) also includes the experimentally obtained indirect exchange energies, $J$, between the Co nanostripe and single Co adatoms as published in Ref.~\cite{Meier_science_2008}. A positive $J$ corresponds to a ferromagnetic coupling while a negative value corresponds to an antiferromagnetic coupling. A damped oscillatory exchange interaction is present in the same range where the exponentially decaying Pt vacuum spin-polarization is measured. It was shown in Ref.~\cite{Meier_science_2008} that the exchange interaction can be described by RKKY~like exchange and follows in a good agreement a 1D range function. In case of a strong contribution of the Pt polarization to the magnetic coupling one would expect a dominance of ferromagnetic or antiferromagnetic coupling for the overall magnetic exchange interaction. Such an effect would become visible by a shift of the RKKY-curve towards positive or negative exchange energies which is not observed.  These observations raise the question, how exactly the measured Pt spin-polarization is linked to the induced magnetization within the Pt surface. 

\section{Theoretical method} 

In order to obtain deeper insight into the relation between the measured spin-polarization in the vacuum and the induced magnetization we performed calculations on three different arrangements of Co \textit{on} or \textit{in} a Pt(111) surface layer as shown in Fig.~\ref{fig:4}~(a). First, we considered a single Co atom deposited on (adatom) and embedded in (inatom) the first layer of Pt(111). These two arrangements differ mainly in the number of next neighboring Pt atoms which is tripled for the inatom with respect to the adatom case. Therefore a comparison of these two cases provides us with important information concerning the hybridization of the Co electronic states with those of the Pt surface leading to the magnetization of the surrounding Pt atoms. 

In order to model the experimental setup as close as possible we constructed a chain of five Co atoms embedded in the surface of Pt(111). This model arrangement reflects the experimental fact that the Pt surface atoms which show a vacuum spin-polarization are located at the same layer than the Co atoms which form the nanostripe. The chain is oriented along a direction perpendicular to the direction probed experimentally concerning the spin-polarization (cp.~Fig.~\ref{fig:3}~(a)). The exact experimental setup is of course difficult to achieve since a non-regular step edge of platinum interfacing a cobalt stripe is impossible to reproduce with  
methods based on Density Functional Theory at the actual stage. 
The method of investigation is the KKR method~\cite{skkr} within the framework of Density Functional Theory. 

KKR is based on multiple--scattering theory. For non--overlapping 
potentials the following angular momentum representation of the Green's function 
$G({\mathbf r}+{\mathbf R}_n,{\mathbf r'}+{\mathbf R}_{n'};E)$ can be
 derived:  
\begin{eqnarray}
\!\!\!\!{G}({\mathbf r}+{\mathbf R}_n,{\mathbf r'}+{\mathbf R}_{n'};E)\!\!\!&=&\!\!\! 
-i \sqrt{E} \sum_{L}{R}_{L}^n({\mathbf r_<};E){H}_{L}^n({\mathbf r_>};E)\delta_{nn'}\nonumber\\
\!\!\!\!&+& \sum_{LL'}{R}_{L}^n({\mathbf r};E) {G}_{LL'}^{nn'}(E)
{R}_{L'}^{n'}({\mathbf r'};E)
\label{eq:th1}
\end{eqnarray}

$\mathbf{R}_{n}$, $\mathbf{R}_{n'}$ refer to the 
atomic positions and $E$ is the energy. 
$\mathbf{r_<}$ and $\mathbf{r_>}$ denote the shorter and 
longer of the vectors $\mathbf{r}$ 
and $\mathbf{r'}$ which define the position in the Wigner--Seitz (WS) cell centered around ${\mathbf R}_n$ 
or ${\mathbf R}_n'$. The 
${R}_{L}^n(\mathbf{r};E)$ and ${H}_{L}^n(\mathbf{r};E)$ 
are the regular and irregular solution of the Schr\"odinger 
equation. 

The structural Green functions ${G}_{LL'}^{nn'}(E)$ are then obtained by solving the Dyson equation for each spin direction.
\begin{eqnarray}
\!\!\!\!{G}_{LL'}^{nn'}(E) &=& {g}_{LL'}^{nn'}(E) \nonumber\\
\!\!\!\!&+& \!\!\!\!\!\!\sum_{n'',L''L'''}{g}_{LL''}^{nn''}(E) 
{\Delta t}_{L''L'''}^{n''}(E) {G}_{L'''L'}^{n''n'}(E)
\label{eq:th2}
\end{eqnarray}

The summation in (\ref{eq:th2}) is over all lattice sites $n''$ and angular momenta $L''$, 
$L'''$ for which the perturbation 
${\Delta t}_{L''L'''}^{n''}(E)$ 
between the ${t}$ matrices of the real 
and the reference system is significant. ${g}_{LL'}^{nn'}$ is  
the structural Green function of the reference system, {\it i.e.} in our 
case the ideal Pt(111) surface.

The real-space solution of the Dyson equation requires a cluster of perturbed atomic potentials that include the potential of Co impurities and the first shell of neighboring cells. It is important to note that the vacuum region is filled with cellular (Voronoi) potentials. Since our aim is to explain the STM measured spectra, we use the Tersoff-Hamann theory~\cite{TH1983,TH1985} and calculate the local density of states in the vacuum at ~4.1 \AA~above the substrate.
After obtaining a self-consistent Co potential with its neighboring shell, one additional calculation is performed including Pt atoms as well as their neighboring vacuum cells at ~4.1 \AA~above the substrate along a given direction.

\section{Theoretical results}

\begin{figure}
\includegraphics[width=0.5\columnwidth]{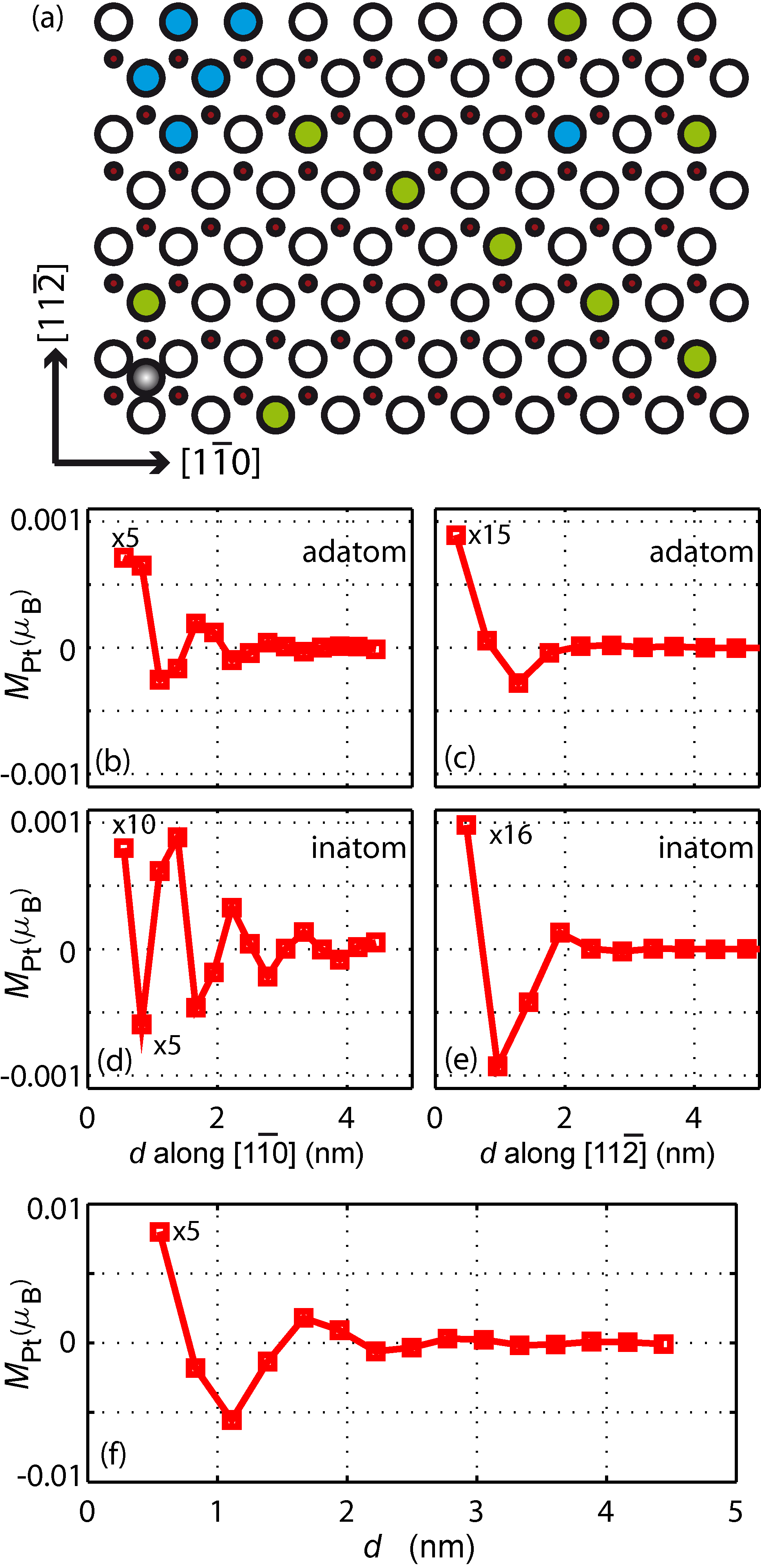}
\caption{~(a)~Sketch of the three considered sample setups for calculations. Large and small open circles represent Pt(111) surface and subsurface atoms, respectively. Filled blue circles indicate the locations of the embedded Co atom chain and the Co inatom. Gray circle marks the position of Co adatom. Filled green circles indicate the closest considered atoms for the calculation of the induced moments in each specific direction. Filled green circles close to a Co atom mark first considered Pt atoms for specific directions.~(b)-(e)~Induced magnetic moments in Pt atoms $M_{Pt}$ for two indicated directions as a function of distance $d$ from a Co adatom and Co inatom.~(f)~Induced magnetic moments in Pt atoms $M_{Pt}$ as a function of distance $d$ from an embedded Co chain for experimentally relevant directions. Some values in (b)-(f) have been scaled down by the indicated factors in order to fit into the figure.\label{fig:4}}
\end{figure}

\begin{figure}
\includegraphics[width=0.7\columnwidth]{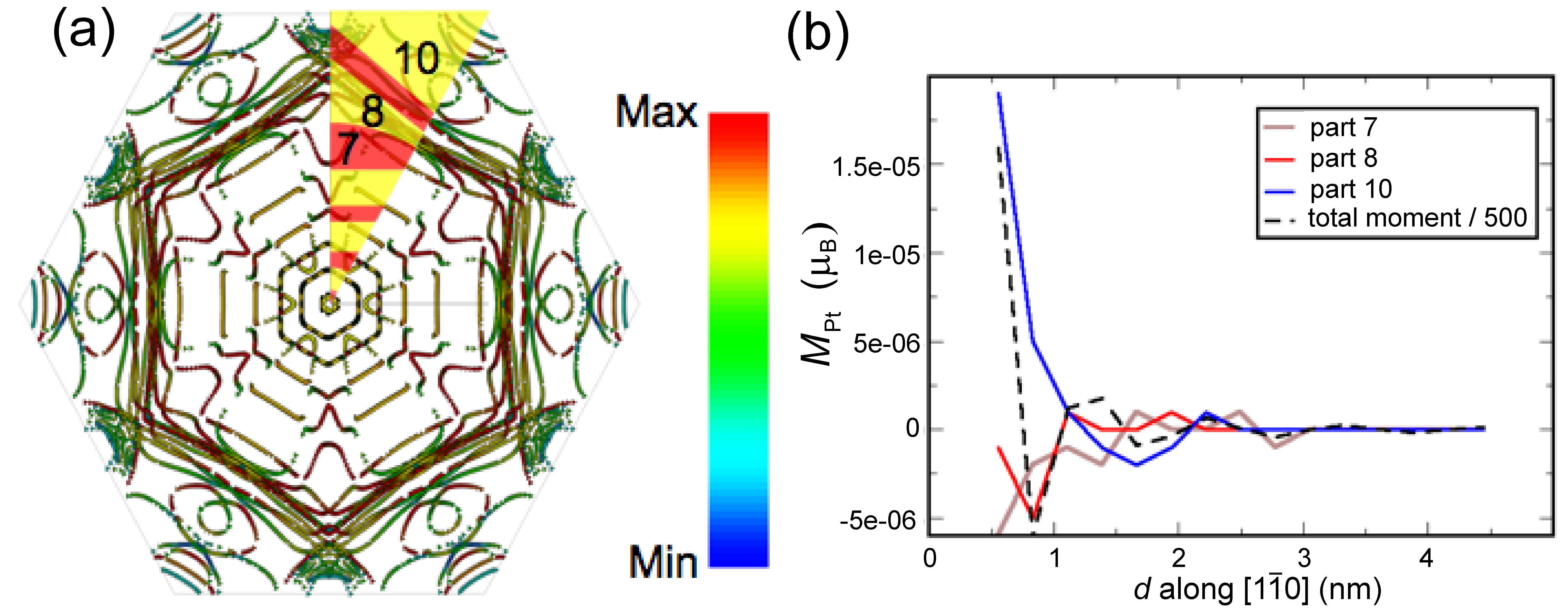}
\caption{~(a) Constant energy contours calculated at the Fermi energy, E$_F$,  where colors represent the degree of localization of the different electronic states on the surface layer of Pt(111). In addition, a triangle divided in 
ten regions is 
superimposed on the energy contours. Depending on the region considered, the induced magnetic moments in the surrounding Pt surface atoms changes. As an example, we plot in (b) the induced magnetic moments of Pt surface atoms along the $[1 \bar{1} 0]$ direction for the inatom case induced by the most contributing constant energy contours: regions 7, 8 and 10. 
  \label{fig:fermi_Pt}}
\end{figure}

For an individual Co adatom and Co inatom we calculated the induced magnetic moments $M_{Pt}$ in the Pt substrate along two directions as indicated in Fig.~\ref{fig:4}~(a). Figures~\ref{fig:4}~(b)-(e) show $M_{Pt}$ as a function of the distance $d$ from the impurity for the $[1 \bar{1} 0]$ and $[1 1 \bar{2}]$ direction. Concerning the $[1 \bar{1} 0]$ direction we find for both arrangements a long range oscillation in $M_{Pt}$ with a wavelength of about 1~nm for the adatom (cp.~Fig.~\ref{fig:4}~(b)) and a slightly smaller one for the inatom (cp.~Fig.~\ref{fig:4}~(d)). The oscillation indicates that $M_{Pt}$ is either ferromagnetically or antiferromagnetically aligned with the Co impurity dependent on the distance. However, the total integrated net moment of the Pt atoms is positive. Along the $[1 1 \bar{2}]$ direction the oscillatory behavior is much weaker than  the one obtained along the $[1 \bar{1} 0]$ direction for both arrangements (Fig.~\ref{fig:4}~(c),(e)). Here more Pt atoms are coupled ferromagnetically to the Co impurity. This directional dependence proves that the induced magnetization is anisotropic which originates from the non-spherical Fermi surface characterizing this system as found in the directional dependent RKKY interactions between Co adatoms on a Pt(111) surface or in the anisotropic induced charge oscillations caused by Co impurities buried below Cu surfaces. ~\cite{zhou_np_2010,weismann_science_2009} A comparison of $M_{Pt}$ for the same direction shows that for the same distances the intensity is always higher for the embedded atom than for the adatom. This emphasizes the importance of the number of neighboring atoms and indicates a dependence of the coupling between the Co and Pt electronic states on the coordination and environment. 
To favor the coupling to the impurity states, the electronic states controlling the studied 
long ranged magnetization must be localized at the surface.
Constant-energy contours at the Fermi energy are plotted in Fig.~\ref{fig:fermi_Pt}(a) for the simulated Pt(111) surface with their relative localization on the surface layer. The degree of localization is depicted in colors: red for maximum localization, blue for minimum. There is a finite number of contours due to the fact that the surface is simulated with a finite number of Pt layers. The shape of the contours is non-trivial indicating the complexity of the problem. This type of calculations indicate the presence of several states which are resonant-like. To measure the degree of coupling between these states and those of the Co impurity, we decompose the Fermi surface in 10 parts represented within the red-yellow triangle in Fig.~\ref{fig:fermi_Pt}(a). Each part includes more or less localized states. Afterwards, we calculate the induced magnetization at the Fermi energy $E_F$ induced by every part. The structural Green function $g$ of Pt(111) needed in Eq.~\ref{eq:th2} is given as a Fourier transform or integral over the first Brillouin zone. This integration can be done for every region defined in Fig.~\ref{fig:fermi_Pt}(a) leading to values that can be plugged into Eq.~\ref{eq:th1} and Eq.~\ref{eq:th2} to compute the contribution of every region in the magnetization of Pt at $E_F$. For the inatom case, it seems that parts 7, 8 and 10 are contributing most to the induced $M_{Pt}$ (cp. Fig.~\ref{fig:fermi_Pt}(b)). By summing up all parts, we approximately recover the total energy integrated magnetization (cp. Fig.~\ref{fig:4}(d)). We do not expect them to be equal since with the decomposition scheme some scattering events cancel each other and other ``back-scattering'' events are not taken into account properly. This theoretical experience demonstrates the non-trivial link between the induced long range magnetization and the constant energy contours of the substrate, their degree of localization on the surface layers and coupling strength with the impurities. 

\begin{figure}
\includegraphics[width=0.7\columnwidth]{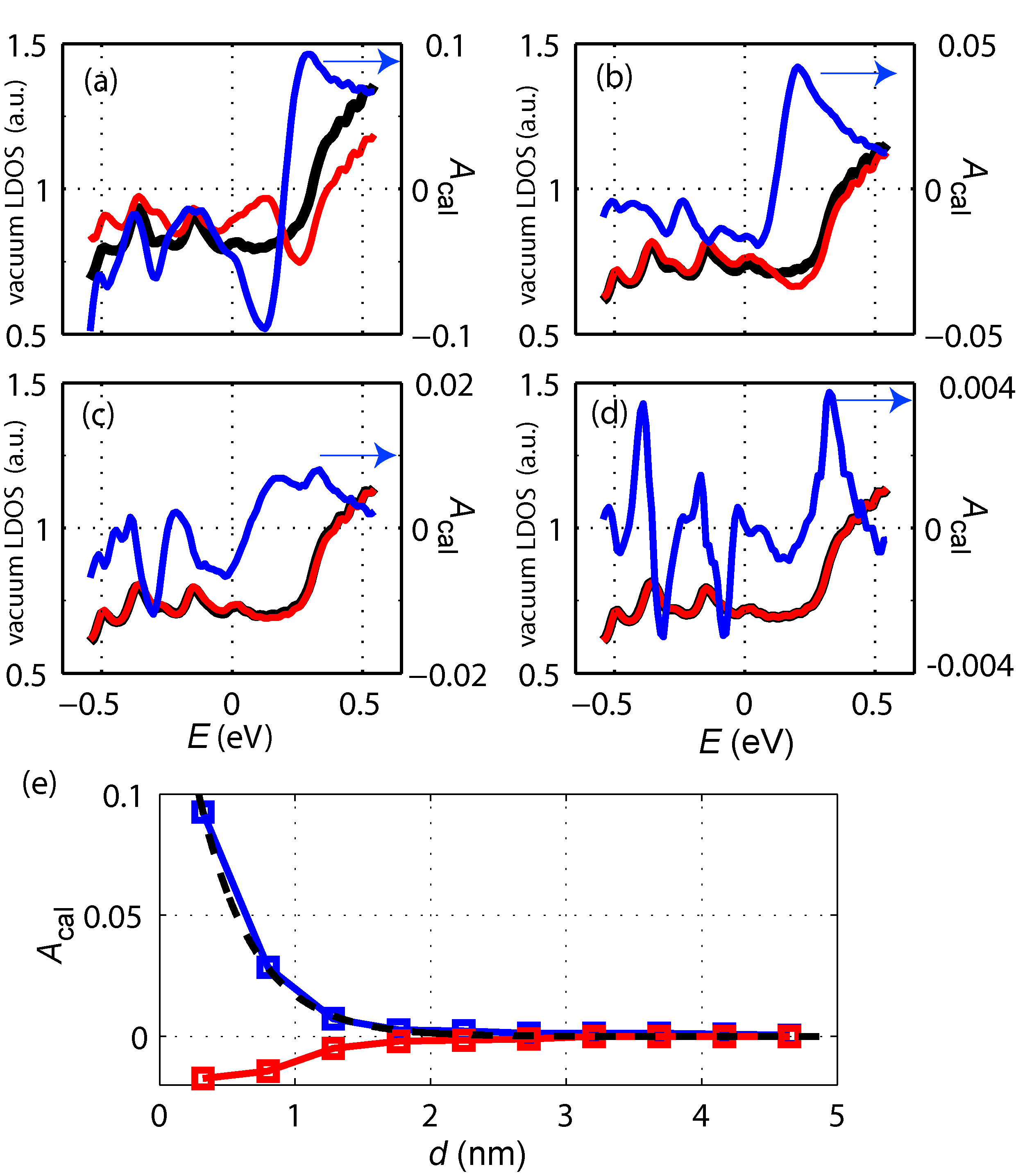}
\caption{~(a)-(d)~Calculated vacuum LDOS above the first, second, third as well as fifth Pt atom close to the embedded Co chain for the majority spin state (black) and the minority spin state (red) as well as the corresponding energy dependent asymmetry (blue). The vacuum LDOS has been calculated at a height of 4.1 \AA~ above the Pt atoms.~(e)~Vacuum asymmetry $A_{cal }$ calculated from vacuum the LDOS above Pt atoms along the direction perpendicular to the chain at +0.3~eV (blue) and -0.1~eV (red). The dashed line shows an exponential fit according to Eq.~\ref{eq:1} to the calculated spin-polarization at 0.3~eV. \label{fig:5}}
\end{figure}

Figure~\ref{fig:4}~(f) shows the $M_{Pt}$ for Pt atoms perpendicular to the embedded Co chain (Fig.~\ref{fig:4}~(a)), as a function of distance $d$ from the chain, which is the setup most similar to the experimental one. In contrast to the experimentally observed decreasing of the vacuum spin-polarization, an oscillating decaying $M_{Pt}$ is observed. Similar to Figs.~\ref{fig:4}~(b)-(e) the curve clearly exhibits the same damped oscillating behavior but shows overall higher intensities which reflects the contributions from all the Co atoms within the chain. 
In order to investigate the relation between $M_{Pt}$ and the energy-dependent spin-polarization we calculated the vacuum LDOS for majority and minority spin states above the Pt atoms along the direction perpendicular to the chain at a vertical distance of 4.1~$\AA$. This corresponds to two interlayer distances from the surface and is the range of the experimental z-height of the tip. 
Figures~\ref{fig:5}~(a)-(d) show the spin-resolved vacuum LDOS for the first, second, third and fifth Pt atom located in the experimental relevant direction. They reveal an intensity increase starting at about +0.3~eV which is due to the Pt surface state.~\cite{wiebe2005} Concerning the difference between the two spin channels it is quite obvious that the Pt atom closest to the chain experiences the strongest imbalance of majority and minority electrons. This is visualized by a corresponding calculated spin asymmetry $A_{\rm cal}(E)$ given by 
\begin{eqnarray}
                    A_{\rm cal}(E) & = & \frac{{\mathrm{LDOS}}_{\rm maj}(E)-{\mathrm{LDOS}}_{\rm min}(E)}{{\mathrm{LDOS}}_{\rm maj}(E)+{\mathrm{LDOS}}_{\rm min}(E)}
\label{average}
\end{eqnarray}
where ${\mathrm{LDOS}}_{\rm maj}(E)$ and ${\mathrm{LDOS}}_{\rm min}(E)$ denote the energy dependent vacuum LDOS for majority and minority electrons. $A_{\rm cal}(E)$ is plotted in Figs.~\ref{fig:5}~(a)-(d) for the Pt atoms as well. These curves reveal that neither the absolute value nor the sign of the spin asymmetry $A_{\rm cal}(E)$ is conserved when scanning at different bias voltages around the Fermi energy. Additionally the absolute value of the spin asymmetry $A_{\rm cal}(E)$ at given energies changes with increasing distance form the Co chain. At some energies even a sign change is observed. Figure~\ref{fig:5}~(e) shows the calculated spin-asymmetry $A_{\rm cal}(E)$ for +0.3~eV and -0.1~eV, which are experimentally relevant, for different distances form the chain. A comparison of these curves with the experimental data obtained at 0.3~V and shown in Fig.~\ref{fig:3}~(b) reveals that the calculated asymmetry $A_{\rm cal}(+0.3eV)$ follows the shape of the experimental curves, i.e. it is always positive and shows an exponentially decaying behavior. A fit as in Eq.~\ref{eq:1} gives a value for the decay length $\lambda$  of about 4~$\AA$ which is less than half of the experimental value. The calculated spin-asymmetry at $-0.1$~eV shows a similar behavior but with reversed sign. This change of sign in comparison to the experiment is most likely due to a change of the tip spin-polarization which is known to occur for a bias voltage range below the Fermi energy.~\cite{zhou_prb_2010}

\section{Discussion}
Recently several theoretical studies concentrated on probing and describing magnetic properties of Co nanostructures on Pt(111) quantitatively and qualitatively. They treated Co in different configurations and environments, like Co overlayers on Pt(111) \cite{sipr_jphyscondsmat2007}, Co nano wires attached to Pt(111) step edges \cite{moscaconte_prb_2008,Baud_PRB_2006} and isolated Co adatoms on bare Pt(111) surfaces \cite{blonski_jop_cond_mat_2009,etz_PRB_2008}. Even though these configurations lead to different coordination numbers, which results in different numbers of underlying Pt atoms per Co atom, they show consistently an induced spin moment $M_{spin}$ of the nearest neighboring Pt atoms in the range from 0.1--0.3~$\mu_{B}$ which is about one magnitude larger than the orbital moments $M_{orb}$. Therefore the total induced magnetic moment $M_{Pt}$ of Pt atoms is mainly determined by the spin moment $M_{spin}$.\\
Additionally it has been found in these calculations that the induced Pt magnetization decreases very rapidly with the distance from the Co structures by about one order of magnitude for the second and third nearest neighbors as shown for the Co nano wires in Ref.~\cite{moscaconte_prb_2008}. Here we probed experimentally and theoretically $M_{Pt}$ for longer distances far from the Co impurities. We find that induced magnetic moments in the surrounding Pt surface atoms are not constantly parallel or antiparallel aligned with the magnetic moment of the Co impurity. The sign as well as the strength of the induced magnetic moments is additionally highly influenced by the strong anisotropy of the Fermi surface of Pt. Both underlines that for the probed arrangements of Co on and in the Pt(111) surface one cannot expect a constantly aligned polarization cloud as found for Co-Pt and Fe-Ir multilayers.~\cite{knepper_prb_2005,Das_JMMM_2010}\\
The apparent contradiction of the measured monotonously decaying $A_{spin}$ in the vacuum and the calculated oscillating $M_{Pt}$ for the embedded Co chain arrangement can be explained by local changes of the electronic structure of the Pt atoms close to the embedded chain (cp. Figs.~\ref{fig:5}~(a)-(d)). It is evident also that the hybridization between the Pt and the Co states changes with increasing the distance from the chain.  
Therefore also the spin-averaged LDOS changes laterally which can be obtained by calculating the arithmetic mean of the LDOS for both spin types in Figs.~\ref{fig:5}~(a)-(d). According to Ref.~\cite{Wortmann_PRL_2001} the measured spin-resolved $dI/dU$ signal and the deduced spin-asymmetry is a measure of the  energy dependent spin-polarization of the sample. {\it This quantity is only a measure for the magnetization, which is an integrated quantity of majority and minority states up to the Fermi energy, if the spin-averaged LDOS is constant.} Therefore the induced magnetization of the Pt cannot be deduced from our experimentally detected vacuum spin-polarization in the Pt only.\\      

\section{Conclusions}
In conclusion, we have performed SP-STM measurements on Pt(111) in the proximity to Co nanostripes at 0.3~K. By probing locally a spin-polarization of Pt, we observed for the first time induced magnetic moments in a non-magnetic material on a local scale. The measured vacuum spin-polarization decays exponentially as a function of the distance from the Co nanostripe with a decay length of about 1~nm.\\
Self-consistent electronic-structure calculations of a Co chain embedded in the Pt(111) surface, of the neighboring Pt atoms and of the vacuum LDOS above the Pt allow us to prove that the measured spin-polarization is induced by an oscillating and highly anisotropic magnetization within the Pt surface in the proximity to Co. By investigating the Fermi surface contours of Pt(111) and their degree of localization on the surface layer, we found several states with anisotropic shapes that could couple to the electronic states of Co impurities and thus contribute to the long range induced magnetization.

\section{Acknowledgments}
We acknowledge financial support from the SFB 668 and
GrK 1286 of the DFG, from the ERC Advanced Grant
"FURORE," and from the Cluster of Excellence "Nanospintronics."
S. L. wishes to thank the Alexander von Humboldt Foundation for a
Feodor Lynen Fellowship and thanks D. L. Mills for discussions and hospitality. The
computations were performed  at the supercomputer JUROPA
at the Forschungszentrum J\"ulich.

\end{document}